\documentclass[epj]{svjour}
\usepackage{lineno,hyperref}
\modulolinenumbers[5]
\usepackage{amsmath}
\usepackage{amssymb}
\usepackage{amstext}
\usepackage{geometry}
\usepackage{units}

\usepackage{comment}
\usepackage{xcolor}

\newcommand{\pbar}{\overline{\mathrm{p}}}
\newcommand{\Hbar}{\overline{\mathrm{H}}}

\usepackage[english]{babel}
\usepackage{graphicx}
\usepackage{color, soul} 
         
\usepackage{float}
\usepackage{braket}

\authorrunning{B. Kolbinger et al.}

\titlerunning{$n$-distribution of $\Hbar$ atoms in a beam}

\begin{document}

\title{Measurement of the Principal Quantum Number 
Distribution  in a Beam of Antihydrogen Atoms}
\author{B.~Kolbinger\inst{1}\thanks{Corresponding author (bernadette.kolbinger@cern.ch).}\thanks{\emph{present address:} CERN, Geneva, Switzerland},
C.~Amsler\inst{1}, 
S.~Arguedas Cuendis\inst{1}\footnotemark[2],
H.~Breuker\inst{2},
A.~Capon\inst{1}, 
G.~Costantini\inst{3,4},
P.~Dupr\'e \inst{2}\thanks{present address: CEA CESTA, 33114 Le Barp, France},
M.~Fleck\inst{5},  
A.~Gligorova\inst{1},  
H.~Higaki\inst{6}, 
Y.~Kanai\inst{7},
V.~Kletzl\inst{1},
N.~Kuroda\inst{5}, 
A.~Lanz\inst{1},
M.~Leali\inst{3,4},
V.~M\"ackel\inst{2}, 
C.~Mal\-bru\-not\inst{8},
V.~Mascagna\inst{4,9},
O.~Massiczek\inst{1},
Y.~Matsuda\inst{5},  
D.J.~Murtagh\inst{1},
Y.~Nagata\inst{10},
A.~Nanda\inst{1}, 
L.~Nowak\inst{8},
B.~Radics\inst{2}\thanks{\emph{present address:} Inst. Particle Physics and Astrophysics, ETH Zurich, Zurich, Switzerland},
C.~Sauerzopf\inst{1}\thanks{\emph{present address:} Data Technology, Vienna, Austria},
M.C.~Simon\inst{1}, 
M.~Tajima\inst{7}, 
H.A.~Torii\inst{5}\thanks{\emph{present address:} School of Science, University of Tokyo, Tokyo 113-0033, Japan},
U.~Uggerh{\o}j\inst{11},
S.~Ulmer\inst{2},  
L.~Venturelli\inst{3,4},
A.~Weiser\inst{1},
M.~Wiesinger\inst{1}\thanks{\emph{present address:} Max Planck Institute for Nuclear Physics, Heidelberg, Germany},
E.~Widmann\inst{1}, 
T.~Wolz\inst{8},
Y.~Yamazaki\inst{2},  
J.~Zmeskal\inst{1}
}                     
%

\institute{Stefan Meyer Institute for Subatomic Physics, Austrian Academy of Sciences, Vienna 1030, Austria \and
Ulmer Fundamental Symmetries Laboratory, RIKEN,  Saitama 351-0198, Japan \and
Dipartimento di Ingegneria dell'In\-formazione, Universit\`a degli Studi di Brescia \and INFN, sez. Pavia, Pavia, Italy \and
Institute of Physics, Graduate School of Arts and Sciences, University of Tokyo, Tokyo 153-8902, Japan \and
Graduate School of Advanced Sciences and Engineering, Hiroshima University, Hiroshima 739-8530, Japan \and
RIKEN Nishina Center for Accelerator-Based Science, Saitama 351-0198,
Japan \and
CERN, Geneva, Switzerland \and
Dipartimento di Scienza e Alta Tecnologia, Universit\`a degli Studi dell'Insubria \and
Department of Physics, Tokyo University of Science, Tokyo 162-8601, Japan \and
Department of Physics and Astronomy, Aarhus University
}
\date{Received: date / Revised version: date}
%
\abstract{
The ASACUSA (Atomic Spectroscopy And Collisions Using Slow Antiprotons) collaboration plans to measure the ground-state hyperfine splitting of antihydrogen in a beam at the CERN Antiproton Decelerator with initial relative precision of $10^{-6}$ or better, to test the fundamental CPT  (combination of charge conjugation, parity transformation and time reversal) symmetry between matter and antimatter. This challenging goal requires a polarised antihydrogen beam with a sufficient number of antihydrogen atoms in the ground state. The first measurement of the quantum state distribution of  antihydrogen atoms in a low magnetic field environment of a few mT is described.  
Furthermore, the data-driven machine learning analysis to identify antihydrogen events is discussed.
%
} 
\maketitle

\section{Introduction}
\label{intro}
\begin{figure*}
\centering
\includegraphics[width = 1.0\textwidth]{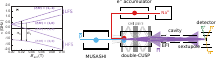}
\caption{Left: Breit-Rabi diagram showing the magnetic field dependence of the four hyperfine states of ground-state antihydrogen, as well as the two accessible transitions in ASACUSA ($\pi_1$ and $\sigma_1$). Right: sketch of the apparatus to measure the hyperfine structure of antihydrogen. Positrons are drawn in red, antiprotons in blue. The synthesised antihydrogen beam is marked in purple. Light purple shows high-field seekers (HFS) and dark purple the low-field seeking (LFS) component of the beam. The grey lines indicate the magnetic field lines of the double-Cusp magnet.}
\label{fig:setup}
\end{figure*} 
The fundamental symmetry of CPT
is a pillar of the Standard Model with no violation observed so far \cite{Tanabashi:2018oca}. As a consequence matter and antimatter are predicted to have equal or sign-opposite intrinsic properties. Nonetheless, an asymmetry between matter and antimatter is observed in the universe \cite{Bennett:2003} and in spite of 
CP violation measurements in mesons~\cite{CPKaons,CPBmes1,CPBmes2,CPDmes} and recent indications for CP violation in the leptonic sector~\cite{T2K-NeutrinoCP_2020} 
to date quantitative explanations are missing.  
This warrants precise measurements of antimatter properties, such as transition frequencies, to compare with their matter counterparts. 
Theories beyond the Standard Model (such as string theory) allow a violation of the CPT symmetry at some level. In particular, the Standard Model Extension (SME) \cite{Colladay:1997vn,Colladay:1998,Kostelecky:2004} provides a general parameterisation of CPT violation and sensitivity guidelines for ato\-mic spectroscopy measurements and other experimental tests. The case of antihydrogen is discussed in dedicated publications in the minimal \cite{Bluhm:1999vq} and non-minimal \cite{Kostel2015} SME.  

Antihydrogen ($\Hbar$) is the simplest stable atom composed solely of antimatter. Hydrogen, its matter counterpart, is one of the most precisely studied atomic systems. The hydrogen ground state hyperfine splitting (GS-HfS) of $\nu\approx 1.42$ GHz has been measured accurately by maser experiments with an absolute (relative) precision of 2 mHz  ($1.4 \times 10^{-12}$)~\cite{Hellwig1970,Karshenboim2000}. 
Since a maser is currently not applicable to antimatter due to the necessary confinement of atoms in a matter enclosure, ASACUSA proposed an in-beam measurement of the GS-HfS of antihydrogen \cite{WidmannEtAl:2001,Widmann:2013,Malbrunot2017aa} and tested the method using a  beam of polarised hydrogen. This resulted in the most precise in-beam measurement of the hydrogen GS-HfS with a relative precision of $2.7 \times 10^{-9}$~\cite{Diermaier2017aa}. 
Comparing  the hyperfine transition frequency of hydrogen and antihydrogen yields one of the most stringent tests of CPT \cite{Widmann:2018}. 

The data presented in this paper were obtained at CERN's Antiproton Decelerator (AD) \cite{Baird:1996ofk}, where ASACUSA produced the first antihydrogen beam in 2012~\cite{Kuroda2014a}. We plan to measure the antihydrogen GS-HfS at CERN's new Extra Low ENergy Antiproton ring (ELENA) \cite{Chohan:2014bja}, initially with a relative precision of $\Delta \nu/ \nu \approx 10^{-6}$ by using the Rabi resonance method~\cite{Rabi1938a} in an antihydrogen beam. Fig.~\ref{fig:setup} (left) shows the Breit-Rabi diagram~\cite{Breit1931} which describes the behaviour of the four hyperfine states of antihydrogen in a weak external magnetic field. The total angular momentum quantum number $F$ and its projection on the quantisation axis $M_F$ are listed for each of the states. With ASACUSA's setup two transitions, $\sigma_1$ and $\pi_1$, are accessible and they are marked in the figure by arrows. The GS-HfS frequency can then be determined by measuring one of the transitions for several field strengths and extrapolating to zero field. Alternatively, it can be calculated by measuring both transitions at the same field strength~\cite{Juhasz2009,Kolbinger2015}.

The GS-HfS frequency of $\Hbar$ has been measured recently in a magnetic trap with  a relative precision of $\approx 4\times 10^{-4}$~\cite{Ahmadi2017}. However, trapping antihydrogen requires strong inhomogeneous magnetic fields which limits the experimental precision. 
In particular, the $\pi_1$ transition frequency  ($F, M_F$: 1,-1 $\to$ 0,0), 
which is sensitive to CPT violations within the SME framework~\cite{Kostel2015}, is prone to systematic biases due to its sensitivity to field inhomogeneities. 
In ASACUSA the interaction region is a low magnetic field environment. By adequate shielding and correction coils the external magnetic fields can be reduced to $\lesssim 1 \, \mu$T~\cite{Thole:2016} with sufficient uniformity for the spectroscopy experiment. 
Furthermore, the temperature of the $\Hbar$ beam can be relatively high (50~K to 100~K)~\cite{Malbrunot2017aa}, much in contrast to trapping experiments which require very cold antiatoms ($\lesssim 0.5$~K). 
On the other hand beam formation and Rabi spectros\-copy are faced with other losses connected to acceptance of solid angle, velocities and quantum states. Therefore, some hurdles still need to be overcome in order to fully exploit this complementary approach. 

Figure~\ref{fig:setup} (right) shows a sketch of the experimental setup to measure the GS-HfS. Antiprotons ($\pbar$) from the AD are stored in the MUSASHI trap \cite{Kuroda:2012zza}. Positrons (e$^{+}$) are obtained from a $^{22}$Na source and a neon moderator then stored in the positron accumulator~\cite{Imao:2010zz}. 
Together they form antihydrogen in the so-called double-Cusp trap~\cite{Mohri2003,Nagata2017aa}.
The double-Cusp trap consists of a multi ringed electrode trap \cite{Mohri_1998} housed within a magnetic field produced by a pair of superconducting coils in an antihelmholtz configuration. A nested penning trap is formed in a region of strong magnetic field before the first of two cusps (see figure \ref{fig:cusp} below) to mix positrons and antiprotons. The purpose of the cusped field is to focus and polarise cold ground state antihydrogen atoms, this is discussed in detail elsewhere \cite{Nagata2017aa}. In this work, it is expected that the antihydrogen formed will be hot hence the focusing and polarisation effect of this configuration will be mini\-mal \cite{Lundmark:2015}.
Details follow in Sect. 2.1.

The polarised $\Hbar$ atoms escape the trap and enter the spectrometer consisting of a microwave cavity \cite{Federmann2012,Malbrunot2019} to induce hyperfine transitions, and a state-analysing sextupole magnet.  In the Rabi-type
resonance method the force from magnetic field gradients exerted on the magnetic moments separates the $\Hbar$ atoms according to their spin states (Stern-Gerlach separation): the sextupole magnet focuses the low-field seeking states and defocusses the high-field seekers. A detector (see Sect.~\ref{sec:det}) records the annihilation signal at the end of the beamline as a function of the microwave frequency applied in the cavity. The challenge lies in producing an intense, focused and polarised source of $\Hbar$ atoms in their ground states. 

This paper is organised as follows: Sect.~\ref{sec:exp} describes the experimental setup and Sect.~\ref{sec:ana} the analysis method for event identification and background rejection in the data recorded by the antihydrogen detector at the end of the beamline. The time distribution of the $\Hbar$ atoms arriving at the detector is described in Sect.~\ref{sec:timedist}. Finally, we  present in Sect.~\ref{sec:qdist} the first measurement of the distribution of the principal quantum number of the $\Hbar$ atoms exi\-ting the double-Cusp trap.

\section{Experimental setup}
\label{sec:exp}

\begin{figure*}
\centering
\includegraphics[width=\textwidth]{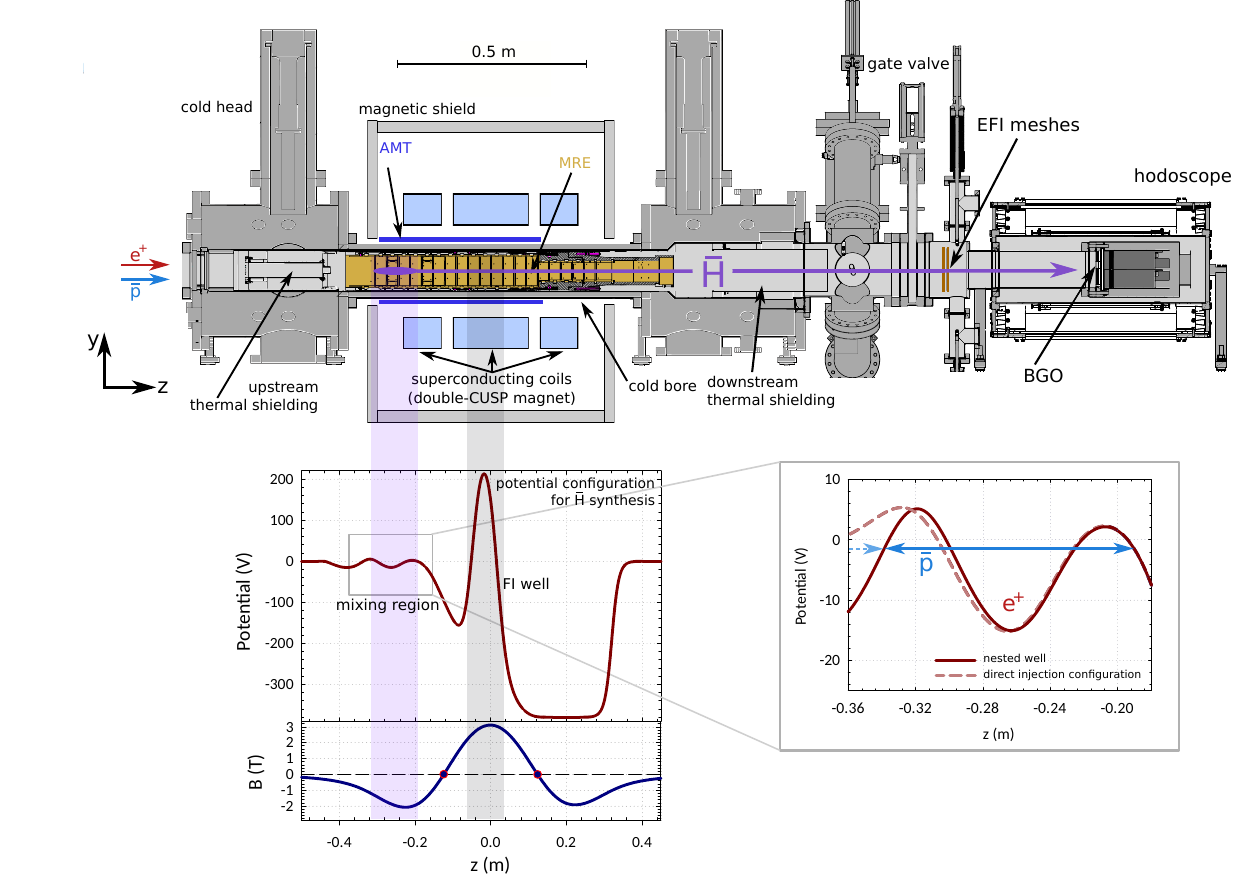}
\caption{Longitudinal cut through of the double-Cusp trap (top), and electric potential (bottom) and on-axis magnetic field configuration (bottom).
 The multi-ring electrode (MRE) is drawn in gold, the superconducting anti-Helmholtz coils in light blue. The $\overline{\text{p}}$ and  $\text{e}^+$ enter from the left and the $\Hbar$ atoms exit to the right. The AMT detector around the mixing region is shown in dark blue. Mixing and field ionisation regions inside the Cusp (FI) are marked by the purple and grey shaded areas. The inset (rectangle) on the right shows the potential configurations used for the direct-injection mixing method \cite{Kuroda2014a}. The dashed blue arrow symbolises the injection of $\overline{\text{p}}$s, the solid blue line the $\overline{\text{p}}$s trapped in the nested well. The downstream beamline with the EFI chamber and the detector is also depicted.}
\label{fig:cusp}
\end{figure*} 

To measure the quantum state composition in the antihydrogen beam we had to modify the apparatus foreseen for the HfS measurement. The $\Hbar$ beamline was shortened to increase the solid angle acceptance at the detector (Sect.~\ref{sec:prod}). The microwave cavity and sextupole magnet were removed and the antiproton annihilation detector was installed after the mixing trap (see Fig.~\ref{fig:cusp} below). The distance between mixing region and 
the centre of the detector was $\approx$185~cm, corresponding to a solid angle of $\approx$0.015\% $\times 4 \pi$. 
An external field ioniser (EFI)~\cite{Sauerzopf2016pro} was inserted between the mixing trap and the detector to deduce information on the principal quantum number distribution of the $\Hbar$ atoms emerging from the trap. Its distance to the production region in the double-Cusp trap is 140~cm where the residual magnetic field of the trap was measured to be $\approx$4 mT.
The EFI consisted of two parallel copper mesh-electrodes, perpendicular to the beam direction (see Fig.~\ref{fig:cusp}). The nominal value of the mesh distance was measured before closing the vacuum chamber to be $\geq 10.0$~mm and $\leq 10.5$~mm everywhere and is therefore assumed to be $10.25 \pm 0.25$~mm.

The voltage polarity was chosen such that antiprotons resulting from the ionisation were deflected in the upstream direction.
The quadra\-tic grid pattern of the meshes had a spacing of \unit[3.04]{mm} with a tines thickness of \unit[0.07]{mm} resulting in a total transparency of both meshes of 95\%. 
However, the grid structure allowed for field penetration resulting in a weaker field than what would follow by dividing the voltage difference by the distance between the meshes. We have performed simulations using the finite element software COMSOL and see an average field, that is about \unit[9]{\%} weaker, see Table~\ref{tab:FI}.

\begin{table}
 \caption{EFI voltage settings and resulting averaged electric fields for minimal and maximal distance $d$ between the meshes. }
\label{tab:FI}       
\begin{tabular}{ccccc}
\hline\noalign{\smallskip}
$\Delta U$ & $U_{1}$ &$U_2$ &  \multicolumn{2}{c}{$|E|$ (V/cm) for $d$ } \\
(kV) & (kV) & (kV) & 10.0 mm  & 10.5 mm   \\
\noalign{\smallskip}\hline\noalign{\smallskip}
10   & $+5$  & $-5$   & 9060  & 8629  \\
0.8   & $+0.4$ & $-0.4$  & 725   & 690   \\
0.14   & $+0.07$  & $-0.07$   & 127   & 121   \\
0.04    & $+0.04$  & $0$     & 36.2  & 34.5  \\			
\noalign{\smallskip}\hline
\end{tabular}
\end{table}

The highest electric field ($E\approx$9~kV/cm) applied during our measurements could ionise substates of the $n$-manifolds down to $n=15$. Four different voltage settings were used to evaluate the quantum number distribution. They are listed in Table~\ref{tab:FI}.

\subsection{Production trap and antihydrogen synthesis}
\label{sec:prod}

The double-Cusp trap for mixing antiprotons and positrons consists of a multi-ring electrode (MRE) \cite{Mohri98} and two superconducting pairs of anti-Helm\-holtz  coils~\cite{Mohri2003,Nagata2017aa} (Fig. \ref{fig:cusp}). The latter provide the inhomogeneous magnetic field to polarise the $\Hbar$ atoms leaving the trap~\cite{Nagata2014a,Lundmark:2015} and focus the low-field seekers entering the spectroscopy section of the apparatus.
Fig.~\ref{fig:cusp} (bottom) shows the electric and magnetic field configurations  along the trap axis. In the mixing region positrons and antiprotons overlap in the nested potential well. A field ionising well in the double-Cusp (FI) is located downstream of the production region. Its electric field, as applied for most of the presented results (see Table \ref{tab:runs}), ionises $\Hbar$ atoms in higher Rydberg states~\cite{Kuroda2014a}. The resulting antiprotons are trapped in the well where they can be released later to estimate the number of  $\Hbar$ atoms which were field ionised.

The ASACUSA Micromegas Tracker (AMT) \cite{Radics2015} consists of a layer of scintillators sandwiched between two layers of Micromegas detectors, located between the magnet and the cold bore around the mixing region. The purpose of the AMT is to reconstruct the annihilation vertices in the trap~\cite{Mackel2018} and to distinguish annihilations on the MRE walls from those on the rest gas in the trap. Annihilations on the walls stem mostly from $\Hbar$ atoms which  are not confined by the electric and magnetic fields of the trap. Wall annihilations provide a complementary way to monitor the $\Hbar$ production process which is independent of quantum states and relies on radial escape.

The production of antihydrogen occurs in 15 to 20 minute cycles. The scheme employed for the presented data is that of direct injection~\cite{Kuroda2014a} of antiprotons
from the MUSASHI trap into a cloud of positrons
in the double-Cusp trap, the method that was  used first to produce antihydrogen in ASACUSA \cite{Enomoto:2010}. 
Several stacks of positrons are collected and then transported to the mixing trap, where they are stored in the nested well potential. Twenty-five stacks are accumulated in the mixing trap, leading to a positron cloud with a radius of 0.9~mm and a density of $6 \times 10^8$ positrons/cm$^{3}$. 
Typically four antiproton shots from the AD are accumulated and cooled in the MUSASHI trap ~\cite{Kuroda:2012zza}. About $6 \times 10^{5}$ antiprotons are then transferred adiabatically to the double-Cusp trap with an energy of 1.5~eV. They are directly injected into the positron cloud with a kinetic energy close to the potential energy of the $\text{e}^+$ plasma, and a narrow energy spread of $0.23 \pm 0.02$~eV~\cite{Tajima_2019}. 

Antihydrogen can be formed via radiative and three-body recombination, where the latter dominates at the temperatures used in our experiment due to its higher cross section~\cite{Gabrielse1988a}. The subsequent evolution of state population is determined by collisional deexcitation and ionisation~\cite{robi2008,Radics2014}.
  
\subsection{Antihydrogen detector}
\label{sec:det}

\begin{figure*}
\centering
\includegraphics[width=0.8\textwidth]{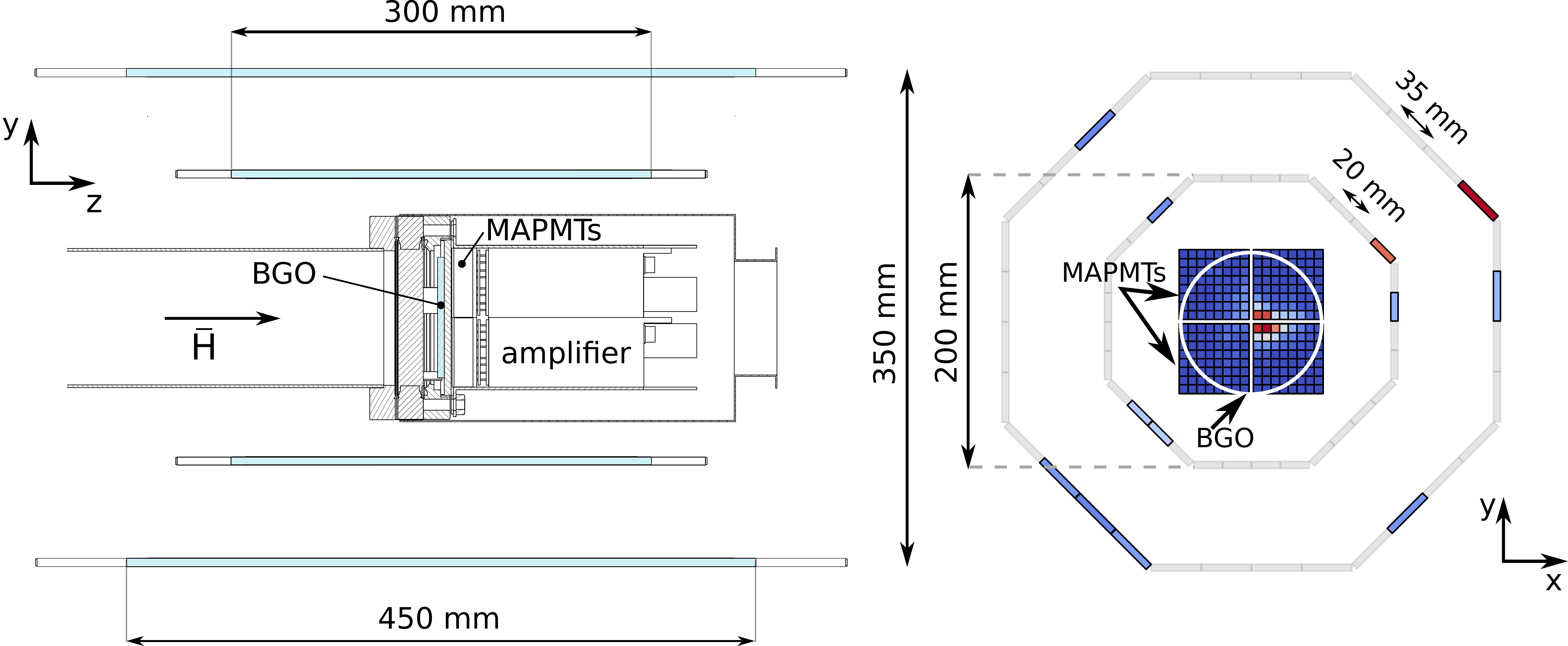}
\caption{Left: side-view of the  $\Hbar$ annihilation detector showing the scintillator bars and BGO crystal in  blue, and the multianode photomultipliers (MAPMT).  Right: $x-y$ cross section of the detector used for online event monitoring, showing a candidate antihydrogen annihilation. The coloured hodoscope bars report channels with  hits. The pixel map in the centre shows the read-out of the MAPMTs. The white circle indicates the position of the BGO crystal.
}
\label{fig:det}
\end{figure*}

The  detector at the end of the beamline is composed of a central BGO (bismuth germanate) crystal 
 which measures the energy deposited by the $\Hbar$ annihilation,  surrounded by a tracking detector to detect the charged annihilation products (mainly pions). A drawing of the detector is shown in Fig.~\ref{fig:det}. 

The BGO crystal is a disk, 90~mm in diameter and 5~mm thick~\cite{nagataBGO,nagataBGO3}, enclosed in a vacuum vessel kept at UHV pressure. The upstream face of the crystal is coated with a 0.7~$\mu$m carbon layer applied by sputter deposition. The carbon layer improves the position resolution of the annihilation point by absorbing the  light reflected back from the downstream face of the crystal. The scintillation light is detected by four Hamamatsu H8500 multianode photomultiplier tubes (MAPMT), placed outside the vacuum and separated from the BGO by a UHV viewport. Each MAPMT is sensitive to an effective area of $49 \times 49$~mm$^2$ (8$\times$8 readout channels). The signals are charge-amplified, digitised and read out by amplifier units mounted directly on the MAPMTs. 
The energy calibration of the BGO crystal is discussed in~\cite{Nagata2018bb}, where details can be found.

The surrounding tracking detector is composed of two layers of each 8 $\times$ 4 plastic scintillator bars (material EJ-200) along the beam direction, arranged octagonally~\cite{Sauerzopf2016bb}. The inner bars are 300~mm long and 20~mm wide, the outer ones 450~mm long and 35~mm wide. Inner and outer bars are 5~mm thick. The dia\-meter of the inner layer is 200~mm, that of the outer one 350~mm. Both layers cover a solid angle of $\approx 80$ \%$\times 4\pi$, seen from the centre of the BGO crystal. The light guides glued on both ends of the bars are connected to  pairs of silicon photomultipliers (KETEK 3350TS SiPM), read out by self-developed front-end electronics \cite{ifes}, 
which provide analogue and digital time-over-threshold signals for all 128 channels. The analogue waveforms are recorded by waveform digitisers (CAEN V1742). The constant fraction time stamps are calculated with the ASACUSA waveform library~\cite{wave}.

The timing information is calculated from the time difference between two bars~\cite{Sauerzopf2016bb}. This helps to distinguish between annihilation events from inside the detector and the external background, such as traversing cosmic rays. The resolution in the time difference between two bars has been measured to be $497 \pm 3$~ps and $551 \pm 5$~ps (full width at half maximum) for the inner and the outer layer, respectively~\cite{Sauerzopf2016bb}. The corresponding hit resolution along the detector axis, given by the time difference between the two ends of the bars, is 59 mm for the inner layer and 73 mm for the outer layer (full width at half maximum)~\cite{KolbingerThesis}.

\section{Event analysis}
\label{sec:ana}

A machine learning analysis was developed to discriminate $\Hbar$ annihilation events from back\-ground and to accurately measure the number of atoms reaching the detector~\cite{KolbingerThesis}. The analysis is based on the supervised method gradient-boosted decision trees\footnote{Library XGBoost~\cite{Chen2016}.} (GBDT) and is trained with measured data. The following sections describe the performed steps of constructing and optimising the event classification analysis: selecting the data set (Sect.~\ref{sec:data}), choosing the discriminating variables (Sect.~\ref{sec:features}) and hyper-para\-meter tuning (Sect.~\ref{sec:parameter_opt}). 

The evaluation of the algorithm is done via several iterations of training and testing. The model is built in the training step with $\frac{2}{3}$ of the events in the data set, and then tested with the remaining events. Based on its response to the test data, the resulting background rejection $\varepsilon_{\text{c}}$ and signal efficiency $\varepsilon_{\overline{\text{p}}}$ are calculated.
Details of the training and evaluation procedure of the algorithm and the  selection of antihydrogen candidate events is described in Sect.~\ref{sec:event_select}. 
The performance of the algorithm is measured via the area under the Receiver-Operating-Characteristics \-(ROC) curve -- the function $\varepsilon_{\text{c}}$ vs. $\varepsilon_{\overline{\text{p}}}$ -- denoted as $A_{\text{ROC}}$ in the following.

\subsection{Data selection and preparation}
\label{sec:data}

The GBDT analysis is trained with measured data consisting of a background and a signal data sample. 
The background is mostly governed by cosmic ray events. The background sample comprises about 3$\times 10^5$  events which have been recorded over a week with beam off. This data sample is free from background occurring during $\Hbar$  data taking which originates from annihilations  upstream of the detector. The trigger rate was $f_{\text{c}} = 0.4687$~Hz. 

Upstream annihilations are not taken into account. These events share all characteristics with annihilations on the detector, apart from their upstream vertex locations (and the lower multiplicity due to the smaller solid angle). 
It was therefore more efficient to  determine the vertex of events after they had been selected by the machine learning analysis.
\begin{figure*}
\centering
\includegraphics[width=0.8\textwidth]{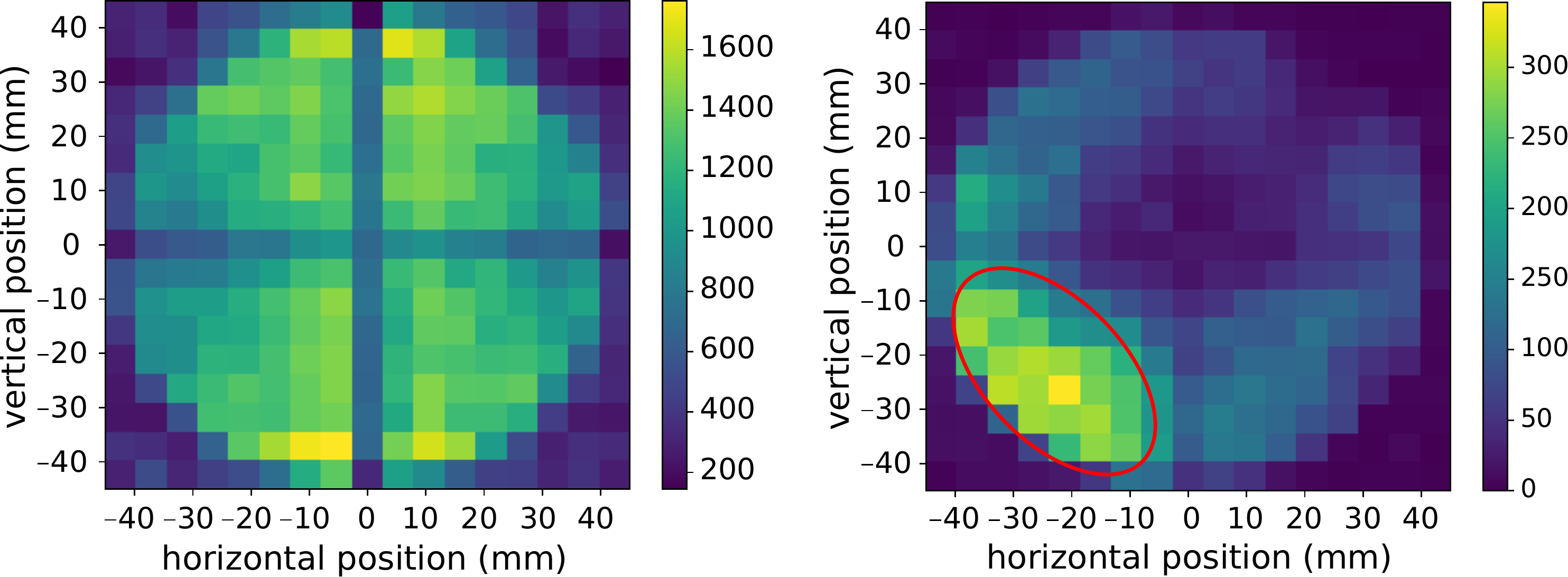}
\caption[]{Two-dimensional histograms of hit positions on the BGO determined by the cluster finding algorithm (for the number of entries in the two-dimensional bins see the colour bar on the right sides of the plots). Left: cosmic ray events. Right: antiproton events. The red ellipse shows the applied cut on the antiproton hit positions for the events included in the signal data sample.}
\label{fig:bgohitpos}
\end{figure*}

Since the antiproton annihilation signature in the ASACUSA detector is identical to the one of antihydrogen atoms and the available number of antihydrogen events is  small, antiproton annihilations are used to train and test the algorithm.
The signal data sample consists of about 6500 antiproton annihilations that have been recorded in dedicated runs. 
Antiprotons of a single AD shot are first trapped and cooled in the MUSASHI trap, before being slowly extracted with 150~eV~\cite{Kuroda:2012zza}. The antiprotons are then transferred through the double-Cusp trap with multi-ring electrodes groun\-ded. The magnetic field of the pair of anti-Helmholtz coils guides the antiprotons towards its exit and the detector. A fraction is defocussed when passing the zero $B$-field regions in the trap and annihilates on the MRE walls.

Due to the possible contamination of the signal data set, careful cuts on  the recorded antiproton data are applied to reduce residual background events.
The following cuts were applied: (1) a cut on the arrival time of events since the slow-extracted antiprotons arrive in a time window of $\approx$4~s length at the detector, (2) events with several spatially separated hits on the BGO are removed due to the possibility of being upstream annihilations i.e. pions from  annihilations on the beam pipe upstream of the detector. Due to the broad and off-centre distribution of annihilation points on the detector, an additional cut (3) on the hit position in the BGO is applied to reduce background contributions. Firstly, a cluster finding algorithm is applied on the two-dimensional $16 \times 16$ pixel read-out of the MAPMTs~\cite{KolbingerThesis}. 
The antiproton hit point on the BGO is  defined as the mean of the pixel positions in the largest cluster, weighted by the energy of the pixels.

Fig.~\ref{fig:bgohitpos} shows two-dimensional histograms of the hit points on the BGO for cosmic rays (left) and for antiprotons (right). While for cosmics the hit positions are uniformly distributed, most antiprotons hit the BGO on the bottom side. 
This is most likely due to a misalignment between the double-Cusp trap with respect to the downstream beamline. 
The red ellipse in the right plot encloses the hit pattern of the  antiprotons and shows the selected cut to reduce the background contribution in the data.

The remaining number of cosmic ray events in the signal sample after cuts has been estimated by considering the total recorded time of antiprotons $t_{\pbar}$ = 372 s (4 s $\times$ 93 runs) and hit area $A_{\text{hit}}$ (the area of the red ellipse in Fig.~\ref{fig:bgohitpos}, right). The residual cosmic contribution was therefore calculated by $f_{\text{c}} t_{\pbar} \frac{A_{\text{hit}}}{A_{\text{BGO}}}$ with the rate of the cosmic events $f_{\text{c}}$ and $A_{\text{BGO}}$ the area covered by the BGO. It amounts to  0.37$\%$ of the total number of events in the signal data sample and is therefore negligible.

It is important to emphasise, that no cuts have been made on event features that will later be used in the machine learning procedure. No cuts have been applied on the cosmic background sample. 

The number of events in the cosmic data set exceeds the number of events in the antiproton data set. Therefore, the balance between the two training sets is ensured by a combination of under- and over-sampling to avoid a biased model.  Initially, the cosmic events of the training sample are randomly under-sampled to 40$\%$ of their total number. Subsequently, the antiproton training data is over-sampled to match the size of the under-sampled cosmic data set. For this purpose, SMOTE (synthetic minority oversampling technique)~\cite{Chawla2002} implemented in the python package imbalanced-learn~\cite{Lemaitre2001} is employed. We emphasise that re-sampling is only applied to the training data. The test data, which are used to deduce cosmic rejection and antiproton efficiency, stay untouched.

\subsubsection{Antihydrogen data preparation}
\label{sec:hbarprep}

During antihydrogen productions runs  the mini\-mal event trigger condition was a hit in the BGO. The antiproton and cosmic data described above were recorded with combined hodoscope and BGO triggers, requiring a hit in the BGO and at least one hit in each hodoscope layer. A hodoscope bar is considered to have a hit if both up- and downstream SiPMs show a signal in coincidence.
To ensure consistency of the input distributions of training and antihydrogen data, the hodoscope trigger condition 
was applied by software to the antihydrogen data after acquisition. The hodoscope trigger is generated by a CAEN V1495 FPGA if one of the inner and one of the outer bars show a digi\-tal signal from the up- as well as from 
the downstream SiPMs. The digital signals are produced with a leading edge discriminator. All 128 analogue signals are recorded by waveform digitisers. 
 The appropriate software cuts on the recorded amplitudes of the hodoscope bars were determined by studying two sets of cosmic data, one with BGO trigger only and one with both detectors in the trigger. Cuts on the first data set were varied and finally chosen such that the event rate of the data set recorded with the combined trigger was reproduced.

\subsection{Discriminating variables and feature selection}
\label{sec:features}
\begin{figure*}
\centering
\includegraphics[width=0.85\textwidth]{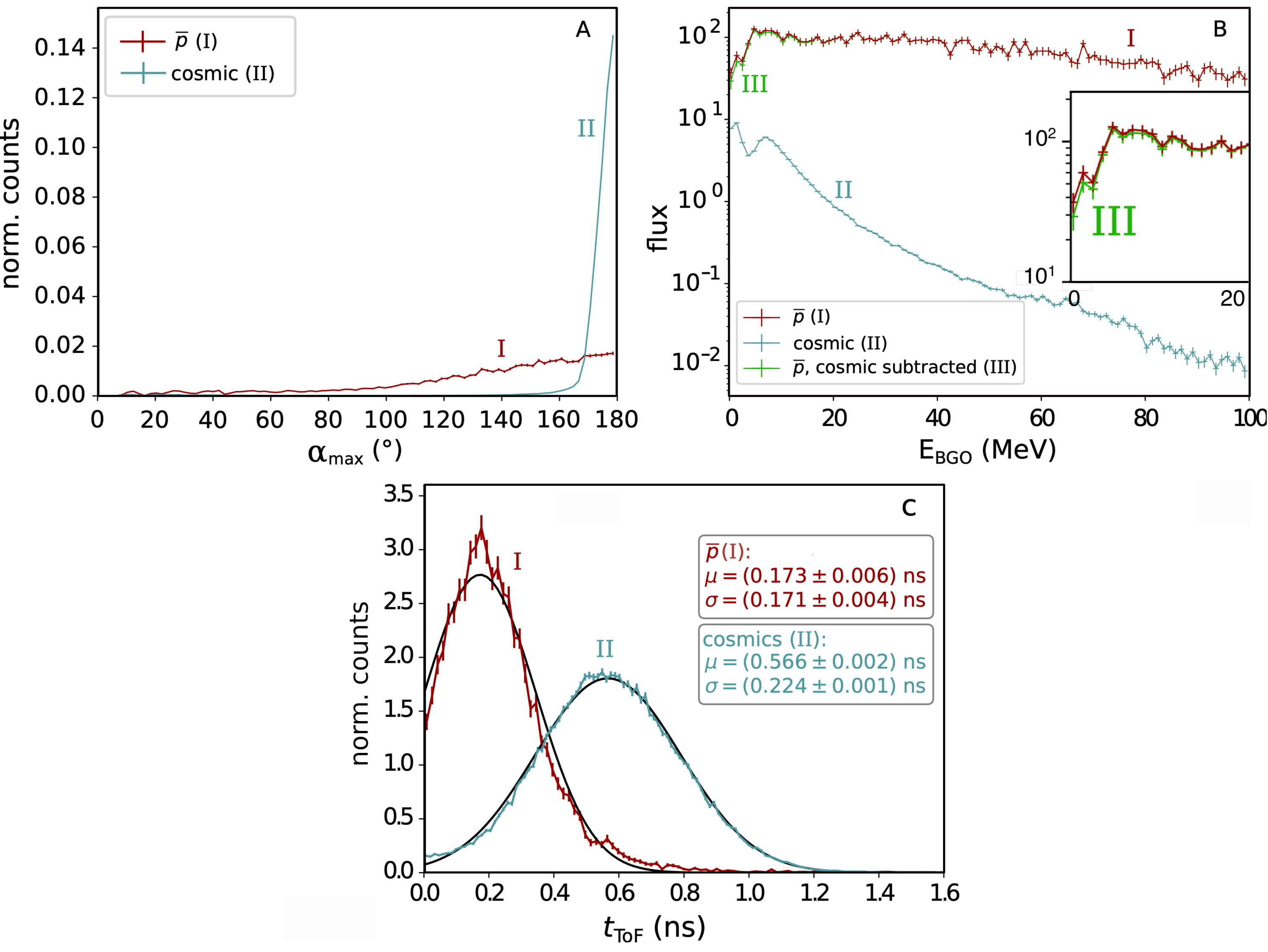}
\caption{Distributions of the three most important features for  antiproton annihilations (I) and cosmics (II). A: $\alpha_{\text{max}}$; B: $E_{\text{BGO}}$, the inset shows the contribution from antiprotons corrected  for cosmic background (III) which affects mainly the 0 -- 20 MeV region; C:  t$_{\text{ToF}}$ with averages and standard deviations for antiprotons and cosmics. The cosmics histogram is scaled to the acquisition time and hit area of the antiproton events. Histograms in A and C are normalised to their area. }
\label{fig:features}
\end{figure*} 
The number and choice of discriminating variables, also called input features in the following, is crucial when using finite data sets, due to the increasing sparsity of the training data with higher dimensionality. 
Twelve potential input features for the analysis have been investigated:
\begin{enumerate}
\item[--] The energy deposit in the central BGO calorimeter $E_{\text{BGO}}$, given by the sum of the output of the MAPMT channels. Details on the energy calibration of the detector can be found in~\cite{nagataBGO3} and \cite{nagataBGO2}.

\item[--] The number of hits in the inner and outer hodoscope ($n_\text{I}$ and $n_\text{O}$). 
\item[--] The number of tracks $N_{\text{tr}}$ in the hodoscope,  which is determined using the $x-y$ information of the bar hodoscope perpendicularly to the beam axis. The algorithm is similar to a Hough transformation and details are described in~\cite{KolbingerThesis}.
\item[--] The time-of-flight $t_{\text{ToF}}$ is given by
\begin{equation}
t_{\text{ToF}} = \sqrt{\frac{1}{N_{\text{tr}}-1} \sum\nolimits_{n=1}^{N_{\text{tr}}} (t_{n} - \overline{t})^2},
\end{equation}
where $t_n = \frac{1}{2}(m_{\text{I},n} + m_{\text{O},n})$ with $m_{\text{I},n}$ and $m_{\text{O},n}$ the average of all mean times of bars belonging to track $n$ in the two layers I and  O, and where $\overline{t}$ is the mean of all $t_n$'s. Annihilation events stem from inside the detector and their annihilation products traverse the hodoscope at approximately the same time, while $t_{\text{ToF}}$ is considerably larger for cosmic rays crossing the detector.
\item[--] The largest angle $\alpha_\text{max}$ between tracks in an event  and the second and third largest angles ($\alpha_\text{2}$ and $\alpha_\text{3}$). 
\item[--] The mean $\alpha_\text{Y}$ of all angles between the tracks and the vertical axis.
\item[--] The number $N_\text{h}$ of horizontal tracks and the number of tracks in the upper and lower hodoscope halves, $N_\text{u}$ and $N_\text{d}$ respectively. A track is classified as being horizontal if it traverses the left and right side bars of the hodoscope.
\end{enumerate}
Some of these features cannot be calculated for all events. For example,   $t_{\text{ToF}}$ and the angles can only be calculated for events with $N_{\text{tr}} >$ 1. Here, the ability of the chosen algorithm to treat missing values in the feature vector becomes important.

In order to find the best combination and number of input features, the SFFS (Sequential Floating Forward Selection)~\cite{featselect} method is employed\footnote{The implementation of the python package MLxtend (Machine Learning extensions)~\cite{Raschka2018} was used.}. SFFS is a wrapper feature selection method that optimises the performance of the  model chosen (GBDT in this case) by exploring different feature sets. SFFS starts from an empty set and with every iteration adds the feature that results in the largest increase of $A_{\text{ROC}}$, until all twelve features have been added. 
The highest $A_{\text{ROC}}$ is achieved with nine features and the following combination of features listed in order of addition by SFFS: $\alpha_{\text{max}}$, $E_{\text{BGO}}$, t$_{\text{ToF}}$, $\alpha_\text{Y}$, $n_\text{O}$, $\alpha_3$, $n_\text{I}$, $N_\text{h}$ and $N_{\text{tr}}$. The features $\alpha_{\text{max}}$, $E_{\text{BGO}}$ and t$_{\text{ToF}}$ have therefore the highest importance for separating signal from background. Their distributions of cosmic rays and antiproton annihilations are compared in Fig.~\ref{fig:features}.

\subsection{Parameter optimisation}
\label{sec:parameter_opt}

Hyper-parameters define the architecture of a model and they need to be fixed prior to trai\-ning. Examples are the number of trees in the ensemble, or the maximum depth of a tree. Those parameters are optimised to achieve the best performance. The set of parameters corresponding to the maximum $A_{\text{ROC}}$ of the multi-dimensional space of hyper-parameters needs to be found, which involves many model evaluations. Depending on the training time and the number of hyper-parameters, conventional methods, such as grid search, can quickly become too time consuming.

\noindent
For the current analysis, the hyper-parameters have been tuned by utilising the sequential mo\-del-based Bayesian optimisation technique  Tree-Parzen Estimator (TPE)~\cite{Bergstra2011,Bergstra2015}. These methods choose the parameter set to be subsequently explored by taking into account past trials and therefore focusing on promising areas of parameter space. Details on TPE and Bayesian optimisation methods can be found in the literature ~\cite{Bergstra2011,Bergstra2015}. 
The set of found hyper-parameters~\cite{KolbingerThesis} is used throughout all training and evaluation steps in the analysis.

\subsection{Evaluation procedure and selection of antihydrogen candidate events} \label{sec:event_select}

The experimental data set containing antiproton and cosmic ray events was randomly split into a training set which comprised $\frac{2}{3}$ of the data, the remainder was kept to test the model built and to measure its response to unknown events. The model returns a prediction score for each event, a number between 0 and 1, where a value close to zero indicates a cosmic-like event and values close to 1 an antiproton-like event. This procedure was repeated a few hundred times with randomly selected training and test sets and the final results determined by averaging over the individual outcomes. 

\begin{figure}
\centering
\includegraphics[width=0.4\textwidth]{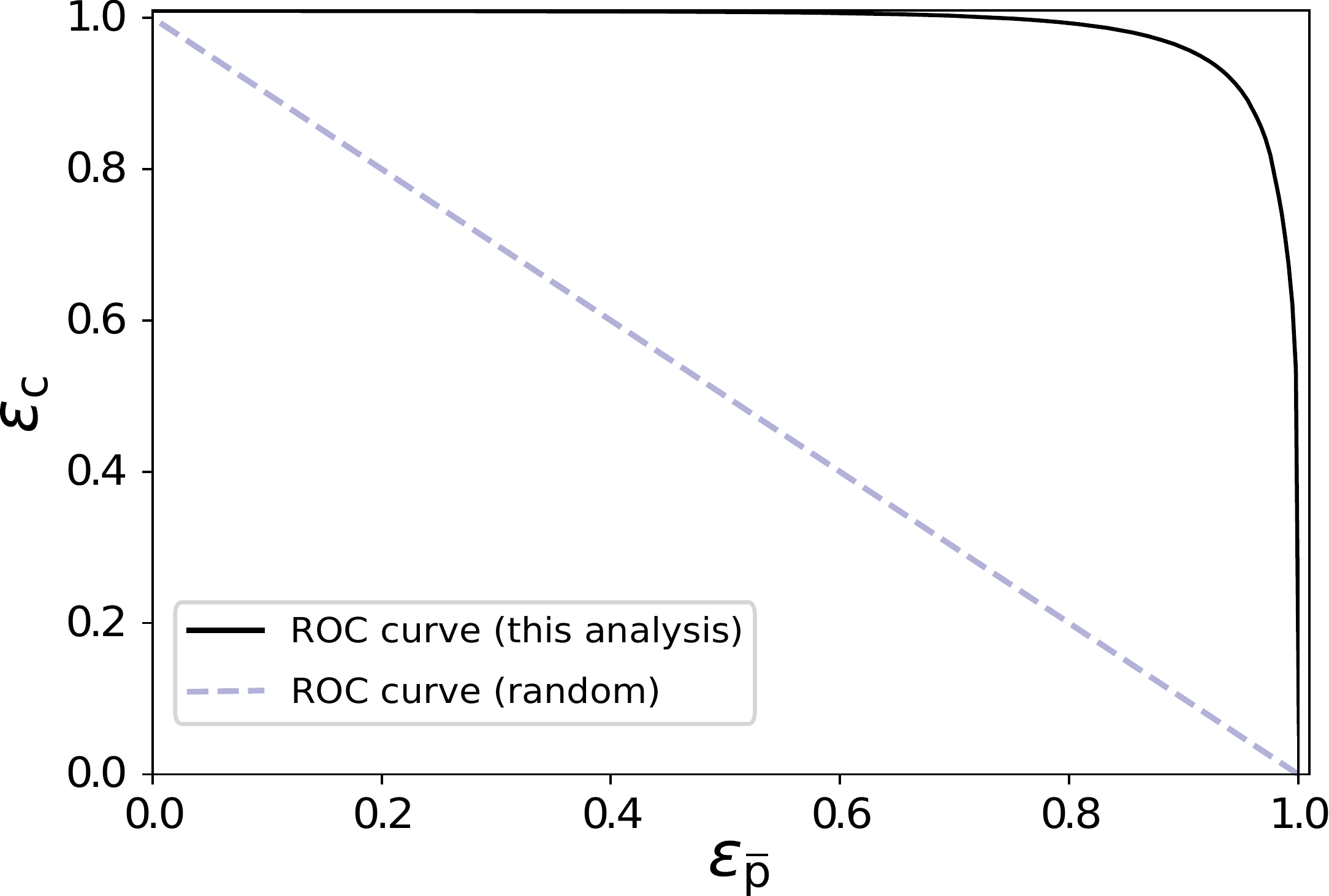}
\caption{The average ROC of our analysis compared to the ROC of a random guessing algorithm.}
\label{fig:roc}
\end{figure}

The average ROC efficiency $\varepsilon_{\text{c}}$ vs. $\varepsilon_{\overline{\text{p}}}$ is displayed in Fig.~\ref{fig:roc}, compared to a random guessing algorithm. 
The average $A_{\text{ROC}}$ of a random guessing algorithm is equal to 0.5, while we obtain $0.9840 \pm 0.0015$ in our analysis~\cite{KolbingerThesis}. Our method has been benchmarked by comparing its results to an ensemble of simple rectangular cuts\footnote{ROOT's TMVA~\cite{tmva} was used.}  which yielded an $A_{\text{ROC}}$ of 0.84, showing that our multivariate analysis leads to a signi\-ficant improvement.

An operating point on the ROC needs to be chosen in order to determine cosmic rejection and antiproton efficiencies. This is done by fixing a cut on the prediction score of events and classifying events with a score below the cut value as background events and those with a score above as annihilation events. 
The antiproton efficiency $\varepsilon_{\overline{\text{p}}}$ and cosmic rejection efficiency $\varepsilon_{\text{c}}$ are then calculated as the ratio of correctly classified events to the total number of events in the test data sets. The false positive rate $f_{\text{fp}}$ is then given by $f_{\text{fp}} = (1 - \varepsilon_{\text{c}}) f_{\text{c}}$, with the rate of cosmic rays equal to $f_{\text{c}} = 0.4687$~Hz. 

The operating point on the ROC curve was chosen by optimising the significance of the $\Hbar$ candidate events found ($N_{\text{obs}}$) with respect to the remaining background ($N_{\text{b}}$) events after analysis. The probability to obtain a larger number $N$ of events than the observed $N_{\text{obs}}$ ($p$-value) is calculated using Poisson statistics~\cite{Behnkea}:
\begin{equation}
\label{eq:pval}
\begin{split}
p = P(N  \geq N_{\text{obs}}|N_{\text{b}}) & = 1 - F_{\text{Poisson}}(N_{\text{obs}}, N_{\text{b}}) \\
& = 1 - \sum^{N_{\text{obs}}}_{k=0} \frac{e^{-N_{\text{b}}}N_{\text{b}}^k}{k!},
\end{split}
\end{equation}
where $F_{\text{Poisson}}$ denotes the cumulative distribution function of the Poisson distribution. The $p$-value can then be translated into the observed significance by $s = F^{-1}_{\text{normal}}(1 - p)$, where $s$ denotes the significance in numbers of $\sigma$ and $F^{-1}_{\text{normal}}$ the inverse cumulative function of the standard normal distribution.

\begin{table*}
 \caption{EFI voltage difference $\Delta U$, status of the field ioniser in the double-Cusp trap, number of mixing runs (\# runs), number of $\Hbar$ candidates ($N_{\text{obs}}$), false positives ($N_b$) within 6~s after mixing start, $n_{\text{threshold}}$ and $n_{100\%}$ (see Section \ref{sec:qdist}).
   }
\label{tab:runs}      
\begin{tabular}{cccccccc}
\hline\noalign{\smallskip}
$\Delta U$ (kV)  & Cusp FI & \# runs  & $N_{\text{obs}}$ & $N_{\text{obs}}$/run &  $N_{\text{b}}$ & $n_{\text{threshold}}$ & $n_{100\%}$  \\
\noalign{\smallskip}\hline\noalign{\smallskip}
10  & ON & 43  &  17  &  0.395  &   $1.979 \pm 0.374$ &  15  & 19  \\
0.8  & ON & 31 &  19  &  0.613  &   $1.427 \pm 0.270$ &  29   & 39  \\
0.14 & ON &16 &  27  &  1.688  &   $0.737 \pm 0.139$ & 46   & 62  \\
0.04  & OFF & 24  &  54  &  2.250  &   $1.105 \pm 0.209$  & $>$60   & $>$60  \\			
\noalign{\smallskip}\hline
\end{tabular}
\end{table*}

By optimising the significance of the antihydrogen candidates found, we obtain $\varepsilon_{\text{c}}$ =  $0.9836 \pm 0.0031$ and $\varepsilon_{\overline{\text{p}}}$ = $0.800 \pm 0.0111$, with the false positive rate $f_{\text{fp}}$ = $0.0077 \pm 0.0015$~Hz. 

The numbers $N_{\text{obs}}$ of $\Hbar$ candidates are summarised in Table~\ref{tab:runs} as a function of field ioniser (EFI) voltage setting. Note that for the lowest voltage setting of the EFI, the FI in the double-Cusp trap was not used. The numbers of runs per EFI voltage setting are listed in column 3 of Table~\ref{tab:runs}. A total of 114 mixing runs were performed to measure the quantum state distribution.  A run was typically 25~s long,  the period from mixing start -- when antiprotons are injected into the positron plasma -- to the time when they are dumped. A total of 159 $\Hbar$ candidates were found, but most antihydrogen candidates however arrive within the first 6~s, see the next section. Table~\ref{tab:runs} lists the 117 candidates arriving within the first 6~s after mixing start. 

A second, independent analysis for signal and background discrimination has been carried out in ASACUSA, see Ref.~\cite{Nagata2018bb} and it is based on two dimensional cuts on event features and does not employ machine learning. A similar detection efficiency of $\tilde{\varepsilon}_{\overline{\text{p}}} = 81\%$ is achieved, the false positive rate is however almost twice as large with 0.012~Hz which demonstrates the effectiveness of a data-driven multivariate analysis.
Furthermore, the analysis~\cite{Nagata2018bb} is based on Geant4 Monte-Carlo simulations. Since discrepancies of multiplicities and energy deposit between Geant4 simulation results and measured data for low energy antiproton annihilations has been observed previously~\cite{aegisEdep,aegisMult1}, the two analyses are not directly comparable. 

\begin{figure*}
\centering
\includegraphics[width=\textwidth]{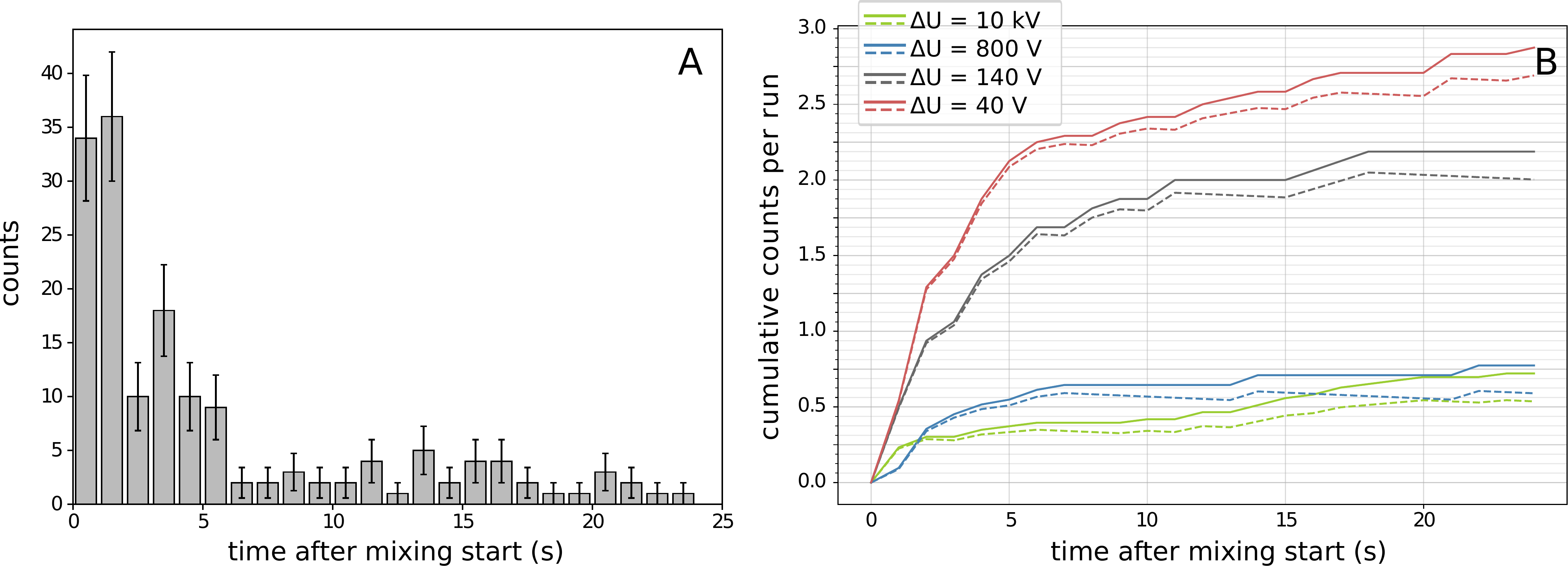}
\caption{A: Distribution of the arrival time of $\Hbar$ at the detector after mixing start. The bin width is 1~s and the error bars are Poisson errors.  B: Cumulative arrival time distribution for the four different EFI settings (colour-coded). The solid lines refer to the total counts per run. The cosmics corrected counts are shown by dashed lines.}
\label{fig:t_dist}
\end{figure*} 

\section{Time distribution}
\label{sec:timedist}

The arrival time of the $\Hbar$ atoms at the detector gives valuable insight into the  $\Hbar$ production process. Fig.~\ref{fig:t_dist}A shows the arrival time distribution of the 159 $\Hbar$ candidates after mixing start. Most events are detected within the first few seconds, 74\% arriving at the detector in the first 6~s. This is most likely due to an axial separation of the antiproton and positron plasmas. 

Fig.~\ref{fig:t_dist}B shows the time behaviour of the cumulative counts for the data sets recorded with different EFI settings. The continuous lines show the total cumulative counts and the dashed lines the cosmic background corrected counts. The steepest increase is always observed within the first few seconds after which the dashed line becomes flat, indicating a negligible number of $\Hbar$ reaching the detector.  

\begin{figure}
\centering
\includegraphics[width=0.45\textwidth]{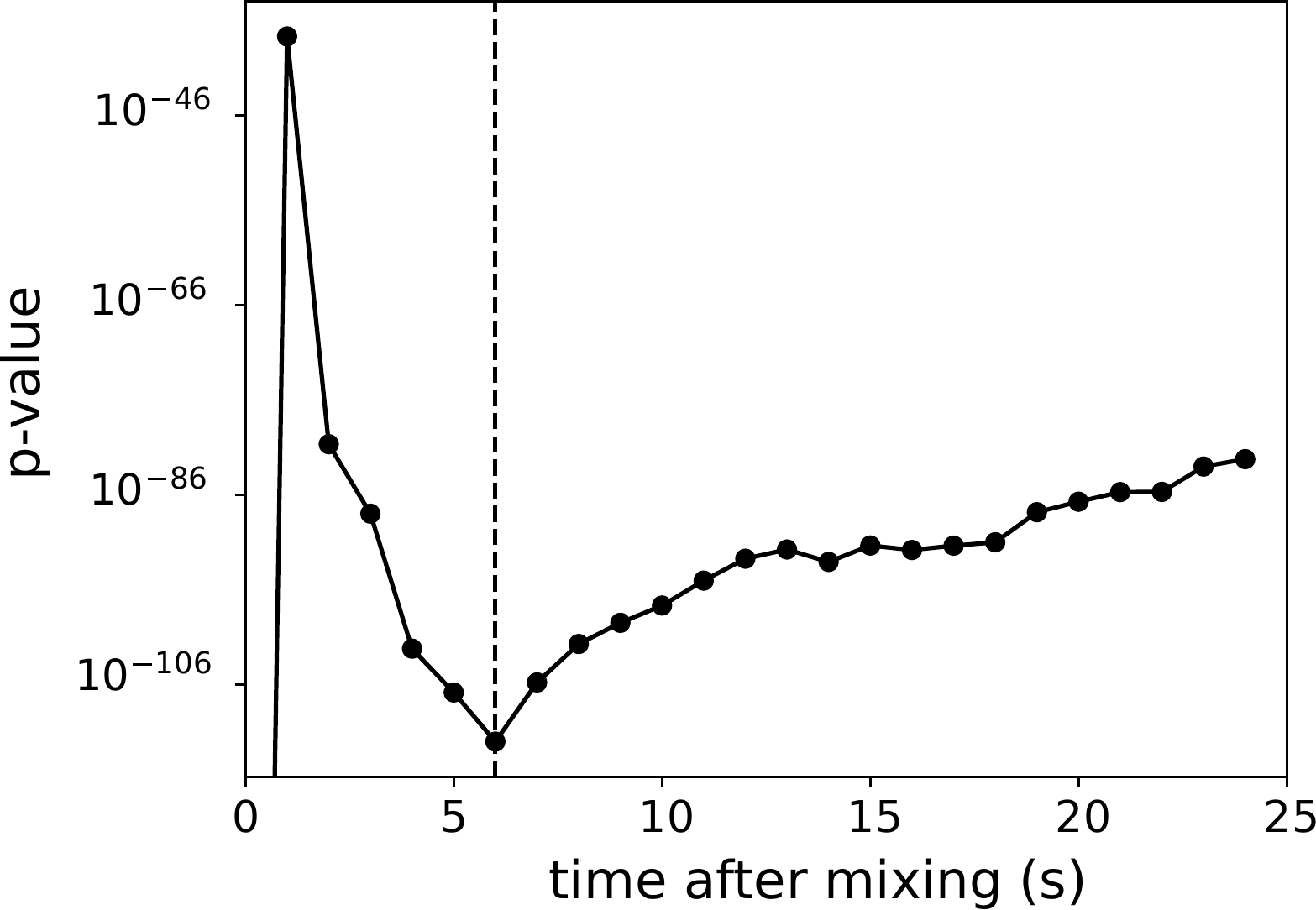}
\caption{$p$-value of $\Hbar$ candidates (all EFI runs) vs. time after mixing. The lowest $p$-value at 6~s is marked by the vertical dashed line.}
\label{fig:t_pval}
\end{figure} 

For further analysis a cut on the arrival time is applied, motivated by the observed steep drop of $\Hbar$ production after a few seconds. The $p$-value of the observed $\Hbar$ counts is calculated via Equation~\ref{eq:pval} for various time intervals. The smallest $p$-value is found with a time interval of 6~s after mixing start (Fig.~\ref{fig:t_pval}).

\section{Quantum state distribution}
\label{sec:qdist}

\begin{figure*}[h]
\centering
\includegraphics[width=0.8\textwidth]{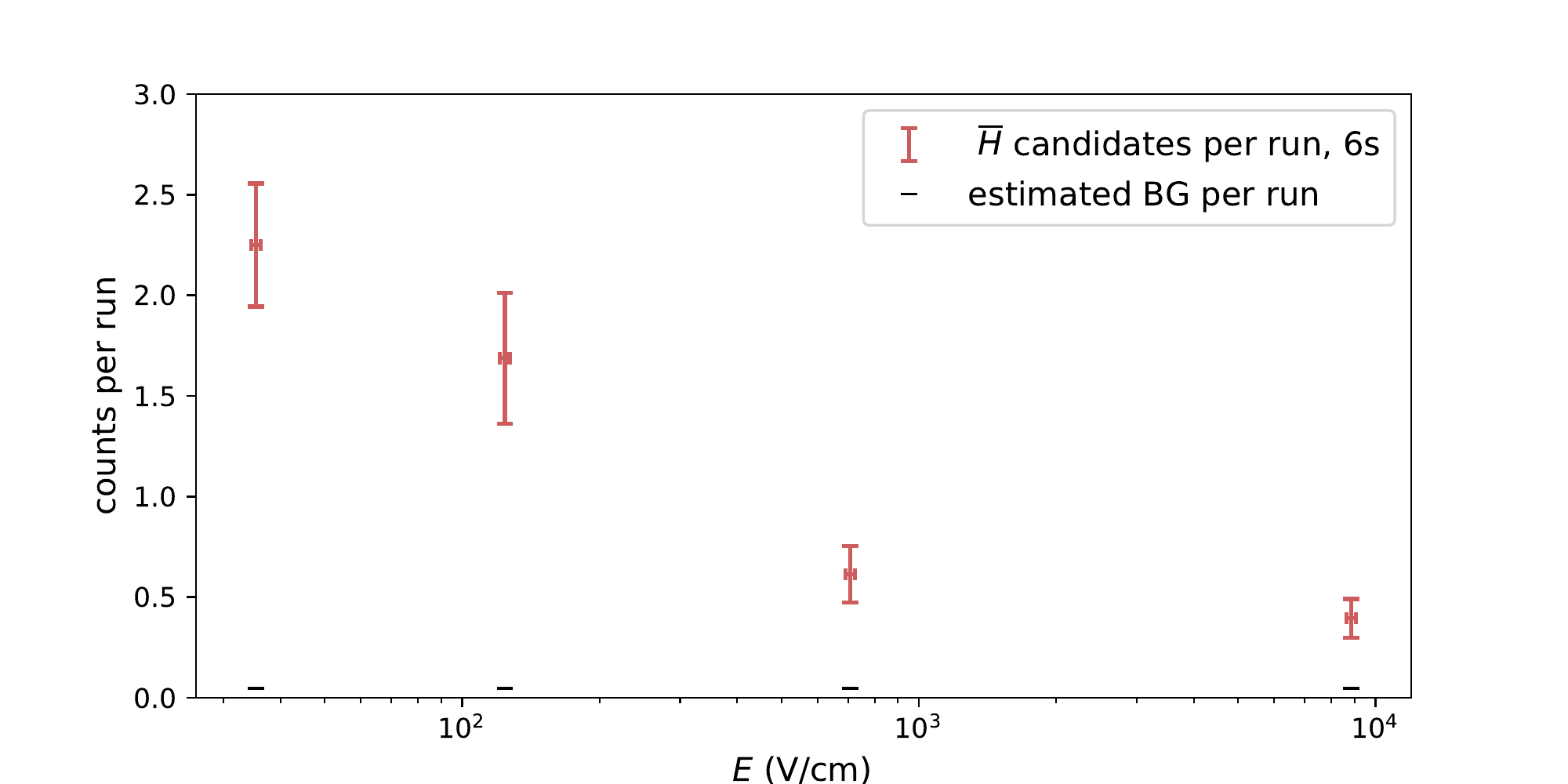}
\caption{Number of $\Hbar$ candidate events per run for the four field ioniser settings listed in Table~\ref{tab:FI} (red crosses).
The mean value of electric fields for $d = 10.25$~mm is plotted as the $x$-coordinate of the crosses. The boundaries of the $x$-error bars correspond to $d=\unit[10]{mm}$ and $d = \unit[10.5]{mm}$.
$y$-error bars show Poisson errors. The estimated background per run is displayed with horizontal, black bars for each of the four field ioniser settings.}
\label{fig:q_dist}
\end{figure*} 

The number of $\Hbar$ candidate events per field ioniser setting as well as the estimated small cosmic background  is plotted in Fig.~\ref{fig:q_dist}.
The distribution of recorded counts at the detector as a function of the electric field at the field ioniser can be used to extract some information on the quantum state distribution of the antihydrogen escaping the formation region toward the detector.
Field ionisation of hydrogen atoms in a static electric field has been discussed in detail in many references (e.g. \cite{Gallagher94} and reference therein).
The classical treatment puts in evidence a saddle point created by the application of the electric field to the purely coulombian potential leading to a electric field threshold beyond which a level is ionised. 
However, the effective potential energy of the Rydberg electron in hydrogen also includes a centrifugal term due to its angular momentum along the quantisation axis \cite{PhysRevA.17.1226} so that for a given excitation energy, states of higher angular momentum will be harder to ionise.
Since we do not know a-priori the sub-level population of the antihydrogen formed within a given manifold, the information given in Fig.~\ref{fig:q_dist} can only be used to provide a range of $n$-manifolds probed for each field ioniser setting.
The quantum mechanical treatment of field ionisation leads to the observation that the lower-lying Stark state (so-called red state) has a maximum electron probability density near the saddle point and thus exhibits large ionisation rates at the ionisation threshold predicted by classical over-the-barrier theory, while the highest-lying state (blue state) has a low density close to the saddle point and thus ionises at higher electric fields. The ionisation threshold for the blue and red states of each $n$-manifold thus gives the minimum electric fields needed to, respectively, start ionising and fully ionise the given $n$-mani\-fold. 

The quantum calculations do not predict an ionisation threshold but provide instead a field-ionisation rate. 
The probability of observing an ion from a given state  depends on the ionisation rate and the transit time through the field (i.e. the $\Hbar$ velocity). The targeted tempe\-rature in this experiment is \unit[50]{K} but there is, as of yet, no measurement of the temperature of the antihydrogen formed with this scheme. Based on simulation \cite{Jonsell_2019} we assume the minimum kinetic energy of the antihydrogen atoms formed to be around \unit[0.001]{eV} but studied the impact on the quantum number addressed if the kinetic energy was up to four orders of magnitude higher. For ionisation occurring while atoms traverse the $\sim\unit[10]{mm}$ field ioniser, this range of energies corresponds approximately to a range $\sim\unit[10^5]{s^{-1}}- \unit[10^7]{s^{-1}}$ in ionisation rate.

\begin{figure*}[h]
\centering
\includegraphics[width=1.0\textwidth]{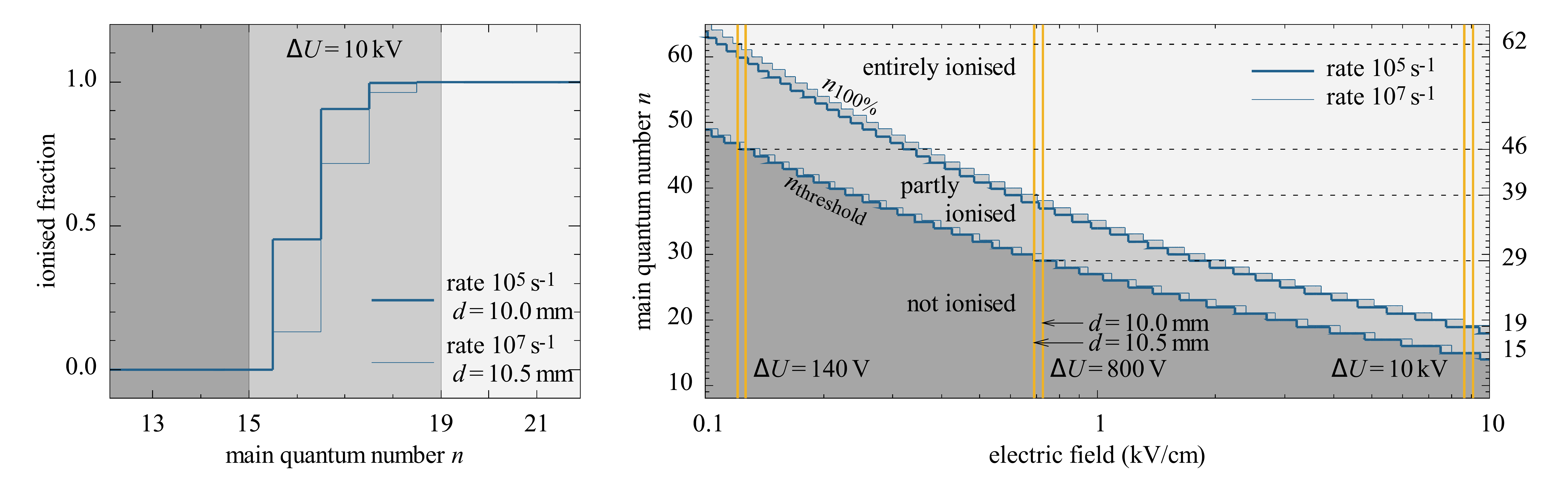}
\caption{Left: Fraction of the states of an $n$-manifold ionised at an electric field of $\sim\unit[9]{kV/cm}$ ($\Delta U=\unit[10]{kV}$) for the two extreme cases of an ionisation rate of \unit[10$^5$]{s$^{-1}$} at $d=\unit[10.0]{mm}$ (thick line) and \unit[10$^7$]{s$^{-1}$} at $d=\unit[10.5]{mm}$ (thin line) calculated using the asymptotic semi-empirical ionisation rates formula provided in \cite{Damburg_1979}. Right: $n_{\text{threshold}}$ and $n_{100\%}$ for ionisation rates of \unit[10$^5$]{s$^{-1}$} (thick line) and \unit[10$^7$]{s$^{-1}$} (thin line). For three voltage settings on the EFI vertical lines mark the range of the resulting electric fields. In both figures three regions indicate in different shades of grey, where $n$-manifolds are not, partly, or entirely ionised.}
\label{fig:n-vs-E-FI}
\end{figure*}

We used the asymptotic semi-empirical ionisation rates formula provided in \cite{Damburg_1979}  to calculate the level width of all sub-states for the $n$-manifolds between $10<n<65$ since our lowest field ioniser setting ($\sim\unit[40]{V/cm}$) starts probing the region $n\sim60$. This formula makes use of the energy of each level calculated by perturbation expansion up to fourth order \cite{bordas_1993} and was checked against exact numerical calculations \cite{Damburg_1976} for different sub-levels up to $n=30$.
However, the residual magnetic field at the location of the field ioniser which is mostly parallel to the electric field and of magnitude $B\sim\unit[4]{mT}$, which we thus considered negligible for low $n$, will start playing a role for high $n$-states. Effects from non-adiabatic coupling of high $n$-states will also affect the ionisation rates~\cite{PhysRevA.23.1127}.
Thus, the formula we used will only give an approximate range of ionisation for high $n$-states.
We provide in Fig.~\ref{fig:n-vs-E-FI} (left) an example of the fraction of states of the different $n$-manifolds which would be ionised in a $\sim\unit[9]{kV/cm}$ field at the rates of $\unit[10^5]{s^{-1}}$ and $\unit[10^7]{s^{-1}}$.

As mentioned already, since we do not know the antihydrogen sub-level distribution, we use such plots to provide for a given electric field the highest state which can pass the field ioniser region without being ionised ($n_{\text{threshold}}$) and the lowest state which is fully ionised ($n_{100\%}$).
We provide in Fig.~\ref{fig:n-vs-E-FI} (right) the $n_{\text{threshold}}$ and $n_{100\%}$ as a function of the electric field up to $n=65$.
The uncertainty on the electric field strength, for the parameters considered, contributes similarly to the uncertainty in $n_{\text{threshold}}$ and $n_{100\%}$ as the range in ionisation rates.

 Fig.~\ref{fig:n-vs-E-FI} shows that the highest field ioniser setting ($\sim\unit[9]{kV/cm}$) can ionise down to the $n=15$ manifold.

In Table \ref{tab:runs}, we provide the approximate range of $n_{\text{threshold}}$ and $n_{100\%}$ probed for each field ioniser setting stressing again that for high $n$ this range is an indication rather than an accurate statement. The results presented here update the analysis published in \cite{Malbrunot2017aa} and the interpretation of the results which ignored the Stark shift of the Rydberg states and thus assumed a classical field for ionisation of $E=1/16n^4$ to estimate the lowest $n$-manifold probed by each field ioniser setting.   


\section{Conclusions} 
\label{sec:sum}

A beam of antihydrogen atoms was produced by injecting antiprotons into positrons in a nes\-ted Penning trap. To distinguish between annihilation events and background events recorded downstream by the $\Hbar$ detector, a data-driven machine learning analysis was developed. The resulting $\pbar$ efficiency is $0.800 \pm 0.011$ and the cosmic rejection $0.983 \pm 0.003$. The false positive rate amounts to ($0.008 \pm 0.002$)~Hz. 

Information on the quantum state distribution was deduced using a field ioniser placed in a low magnetic field region, 140~cm downstream of the production region.
We found 117 $\Hbar$ candidates in 114 mixing attempts with different field ioniser voltage settings during the first 6~s after mixing start, which is the time interval when most candidate events were observed. 

The distribution of principal quantum numbers of the antihydrogen atoms shows that a higher fraction of atoms reaching the entrance of the spectroscopy beamline are in Rydberg states.

The observed production rate of atoms in the lowest lying states ($n_{\text{threshold}} = 15$ and $n_{100\%} = 19$) which are most interesting for ground state hyperfine spectroscopy, amounts to $0.395 \pm 0.096$ per run with an observed significance of 6.8~$\sigma$.
The average spontaneous decay time to ground state from the $n=19$ state assuming equi-popula\-tion of its sub-states is $\sim\unit[130]{\mu s}$ \cite{wolz:2019}. Assuming an average velocity of approximately 1000~m/s (temperature of \unit[50]{K}) -- which is the acceptance limit of our spectroscopy apparatus -- 
the antiatoms are likely to decay to the ground state before entering the microwave cavity (except for antiatoms decaying to the metastable 2s state), which will be installed at the current position of the detector (i.e. \unit[45]{cm} away from the EFI) in the full spectroscopy setup.

 This is the first evaluation of the quantum state distribution in an antihydrogen beam, in a low magnetic field region and down to quantum numbers as low as $n = 15$. The $n$-distri\-bution of high Rydberg states was measured earlier at the AD~\cite{Pohl_2006} but in the presence of a strong $B$-field of 5.4~T where additional complications in the interpretation of the data arise \cite{Driscoll_2004,Vrinceanu_2004,Pohl_2006}.

The results presented here highlight the required steps towards antiatomic beam spectros\-copy, namely increasing the production rate and the population of the lower quantum states.
Estimates on the required production rate necessary to reach the goal of one ppm precision on the ground state hyperfine transition frequency can be made by using our results on hydrogen~\cite{Diermaier2017aa}, which indicate the need of $\approx 8000$ $\Hbar$ for one measurement. 
Therefore, the rate of $\Hbar$ atoms in the lowest quantum states needs to be increased by one to two orders of magnitude.
Hence current efforts focus on investigating methods to boost the production rate, e.g. by decreasing the positron temperature \cite{Radics2014} and also pursuing deexcitation techniques like collisional deexcitation in plasmas \cite{Radics:2016} and light-stimulated deexcitation \cite{wolz:2019} to increase the number of $\Hbar$ atoms in the ground state that reach the cavity.

\section*{Acknowledgements}
The authors are grateful to Eric Hunter for pointing out that the classical model for field ionization used in our previous work is not valid and for his initial inputs to the proper treatment, and for useful discussions with D. Comparat on the field ionisation of hydrogen.
We also express our gra\-titude towards the AD group of CERN. This work was supported by the European Research Council under European Union's Seventh Framework Programme (FP7/2007-2013) /ERC Grant Agreement (291242), the Austrian Ministry of Science and Research, Austrian Science Fund (FWF): W1252-N27, the Grant-in-Aid for Specially Promoted Research 24000008 of Japa\-nese MEXT, Special Research Projects for Basic Science of RIKEN, Universit\`a di Brescia and Istituto Nazionale di Fisica Nucleare and the European Union's Horizon 2020 research and innovation programme under the Marie Sk\l{}odowska-Curie grant agreement No 721559. 

\section*{Author contribution statement}
M.T., N.K., H.H., Y.N., and Y.K. developed and operated the MUSASHI trap and provided ultraslow antiprotons. M.T., N.K., H.H., Y.N., and Y.K. developed and operated the double Cusp trap and provided antihydrogen atoms. D.J.M. designed, constructed, and developed the positron trap, which was operated by P.D., Y.K., H.H., M.T., and N.K. B.R developed the ASACUSA Micromegas Tracking detector and reconstruction code from which V.Mae and B.R. analyzed the data. M.L., V.Mas., and L.V. contributed to develop the pion-tracking detector, its data acquisition system and the off-line software for the data analysis of the tracking detector. Y.N., N.K., and M.T. developed the BGO annihilation detector.
C.M., C.S., B.K., A.C., M.F., J.Z., O.M., and E.W. conceived, built, and tested the hodoscope detector, C.M. and C.S. developed the trigger and data acquisition for the antihydrogen detector. M.C.S. and C.M. developed the External Field Ionizer.\par

The experiment was coordinated and predominantly operated by N.K., V.Mae., Y.Y., B.K., and M.F. 
In addition M.T., N.K., Y.N., P.D., H.H., Y.Mas., V.Mae., B.K., C.S., M.F., O.M., M.W., S.A.C, and H.B. were involved in the data taking campaign. B.K. performed the\\ machine-learning based data analysis. C.M. and M.C.S. took part in critical discussions of the procedure. B.K. wrote the manuscript (section 5 was drafted by C.M.), which was edited by C.A., together with C.M., M.C.S., D.J.M., E.W., and N.K. The study of the ionisation of Stark states has been carried out by T. W. and C. M, with input from M.C.S. and D.J.M.\par

Y.Y. and partly N.K. participated in the basic design and construction of the cusp system including the Cusp trap and the MUSASHI trap, Y.Y. and N.K. conceived and proposed the Cusp trap for making antihydrogen beam for the ground-state hyperfine splitting measurement, E.W. proposed the method for the ground-state hyperfine splitting measurement with antihydrogen.\par

All authors critically reviewed and approved of the final version of the manuscript.

 \bibliographystyle{epj}
 \bibliography{library_3}
%

\end{document}